\documentclass[sigconf, screen]{acmart}

\AtBeginDocument{%
  }

\setcopyright{acmlicensed}
\copyrightyear{2025}
\acmYear{2025}
\acmDOI{XXXXXXX.XXXXXXX} 

\acmConference[ICHEC '25]{International Conference on Human-Engaged Computing (ICHEC 2025)}{November 21--23, 2025}{Singapore}

\acmISBN{978-1-4503-XXXX-X/2025/11} 

\usepackage{xcolor}
\usepackage{subcaption}
\usepackage{float}
\usepackage{booktabs}     
\usepackage{multirow}     
\usepackage{graphicx}     

\begin{document}

\title{Body Management Information Practices on a Female-dominant Platform}

\author{Na Li}
\authornote{These authors contributed equally to this research.}
\email{nzl5264@psu.edu}
\affiliation{%
    \institution{The Pennsylvania State University}
    \city{University Park}
    \state{Pennsylvania}
    \country{USA}
}

\author{Chuhao Wu}
\authornotemark[1]
\email{chuhaow@clemson.edu}
\affiliation{%
    \institution{Clemson University}
    \city{Clemson}
    \state{South Carolina}
    \country{USA}
}  

\author{Hongyang Zhou}
\email{hz648@cornell.edu}
\affiliation{%
    \institution{Cornell Tech University}
    \city{New York}
    \state{New York}
    \country{USA}
}
\author{Huiran Yi}
\email{huiran@umich.edu}
\affiliation{%
    \institution{University of Michigan}
    \city{Ann Arbor}
    \state{Michigan}
    \country{USA}
}

\author{Xuefei Wang}
\email{wangxuefei@uchicago.edu}
\affiliation{%
    \institution{University of Chicago}
    \city{Chicago}
    \state{Illinois}
    \country{USA}
}

\author{Jie Cai}
\authornote{corresponding author}
\email{jie.cai1@outlook.com}
\affiliation{%
    \institution{Tsinghua University}
    \city{Beijing}
    \country{China}
}

\author{Xinyi Fu}
\email{fuxy@mail.tsinghua.edu.cn}
\affiliation{%
    \institution{Tsinghua University}
    \city{Beijing}
    \country{China}
}

\author{John Carroll}
\email{jmcarroll@psu.edu}
\affiliation{%
    \institution{The Pennsylvania State University}
    \city{University Park}
    \state{Pennsylvania}
    \country{USA}
}

\renewcommand{\shortauthors}{Na Li et al.}

\begin{abstract}
With growing awareness of long-term health and wellness, everyday body management has become a widespread practice. Social media platforms and health-related applications offer abundant information for those pursuing healthier lifestyles and more positive body images. While prior Human-Computer Interaction research has focused extensively on technology-mediated health interventions, the user-initiated practices of browsing and evaluating body management information remain underexplored. In this paper, we study a female-dominant social media platform in China to examine how users seek such information and how it shapes their lifestyle choices. Through semi-structured interviews with 18 users, we identify factors including consumerism, poster popularity, and perceived authenticity that influence decision-making, alongside challenges such as discerning reliable methods and managing body anxiety triggered by social media. We contribute insights into how content and media formats interact to shape users' information evaluation, and we outline design implications for supporting more reliable and healthy engagements with body management information.\\
\textcolor{red}{Preprint accepted at ICHEC 2025}


\end{abstract}





\maketitle

\section{Introduction}

The influence of social networks on people's perceptions of their body images is a topic of considerable interest in social science and behavioral research. Previous studies have highlighted how platforms like Instagram and Facebook can contribute to body dissatisfaction through idealized portrayals of beauty and fitness \cite{fioravanti2022exposure}. However, there is also evidence suggesting that social networks can promote body positivity by promoting diverse and inclusive representations of beauty \cite{cohen2019bodypositivity, fardouly2023can}. Research in human-computer interaction (HCI) has extensively examined how people use health information sites \cite{toscos2006chick} and health monitoring tools for body weight management, focusing primarily on populations with medical concerns such as obesity \cite{li2014friend, wang2017understanding, yang2017eliciting, meyer2021natural} and eating disorders \cite{eikey2016privacy}.
As the awareness of healthy lifestyle and positive body image increases, the search for weight loss, healthy diet options, and relevant information is becoming more common even among individuals without obesity concerns. This trend of improving one's appearance through body management can lead to both negative and positive perceptions of self-image. Importantly, people often seek external guidance on how to better manage their body images, and social media platforms are becoming popular sources of such information, regardless of their professional or scientific credibility. Consequently, social networks can create an information cocoon in which users' beliefs about body image are reinforced by the content they choose to interact with.

In particular, women in male-dominated cultures can be more susceptible to the influence of social media on the perception of their body images. Cultural norms in many Asian societies often emphasize slimness and fair skin as ideal standards of beauty, which are frequently reinforced by social media content. This cultural stereotype, combined with the power of social networks, can intensify the desire to improve body image among women in these societies \cite{jackson2020asian}. Despite the strong influence of traditional beauty standards, recent years have also seen significant growth in feminism and body positivity movements on social networks in these cultures. Several social media platforms in China have become hubs for activists and influencers advocating for a broader and more inclusive definition of beauty \cite{lang2024beauty}. Influencers and celebrities who advocate for body positivity and feminist ideals are gaining a large following, contributing to a gradual shift in public perception. This growing awareness is fostering a more supportive online environment.

The role of social media platforms can often be bidirectional, and research is needed to understand how social media affordances interact with cultural norms and gender discrepancies in determining female users' body improvement behaviors. In the context of RedNote (also known as Xiaohongshu), these dynamics are particularly intriguing due to the platform's unique blend of lifestyle content and community-driven advice. With its large female user base, RedNote serves as a microcosm of these broader social changes, where users negotiate and reconstruct their body image in the context of evolving feminist discourses. Through exploration of the user navigation of body improvement on RedNote, this study seeks to answer two primary research questions:

\begin{itemize}
    \item How does RedNote's influencer culture and commercialization shape users' evaluation of body management information?
    \item What challenges do users face when using RedNote to seek body management information?
\end{itemize}

By addressing these questions, this study aims to shed light on the complex interplay between social media, cultural norms, body image perceptions, and the lifestyle changes of female users. Understanding the relationship is crucial for promoting healthy body management and mental well-being among women in male-dominated cultures. The findings could inform policy recommendations and the design of digital platforms that support diverse and inclusive representations of beauty, ultimately contributing to a more equitable and supportive online environment for all users.

\section{Background and Related Work}
\subsection{Study Context: RedNote}
\begin{figure}
\centering
    \begin{subfigure}{0.2\textwidth}
        \includegraphics[width=\linewidth]{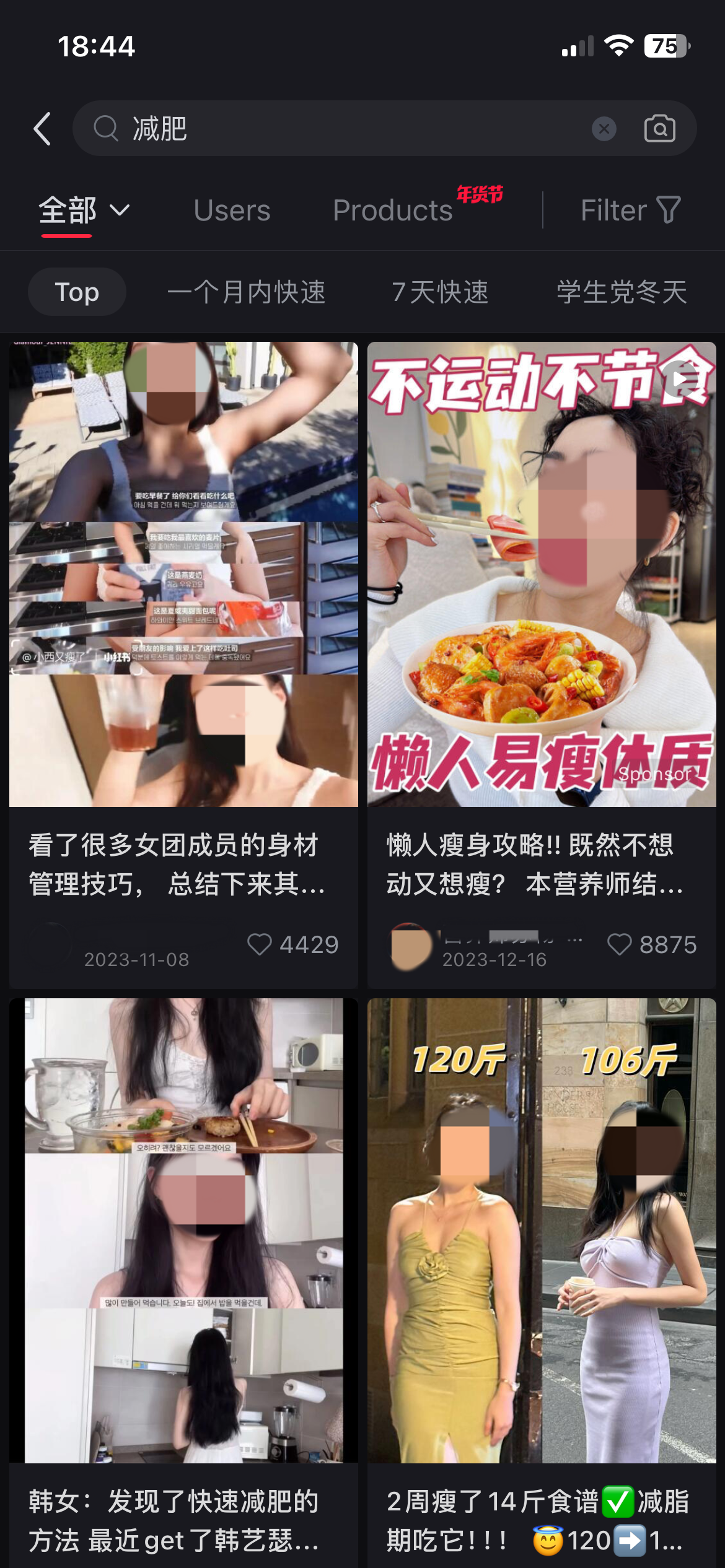} 
        \caption{}
        \label{fig:search}
    \end{subfigure}
    \begin{subfigure}{0.2\textwidth}
        \includegraphics[width=\linewidth]{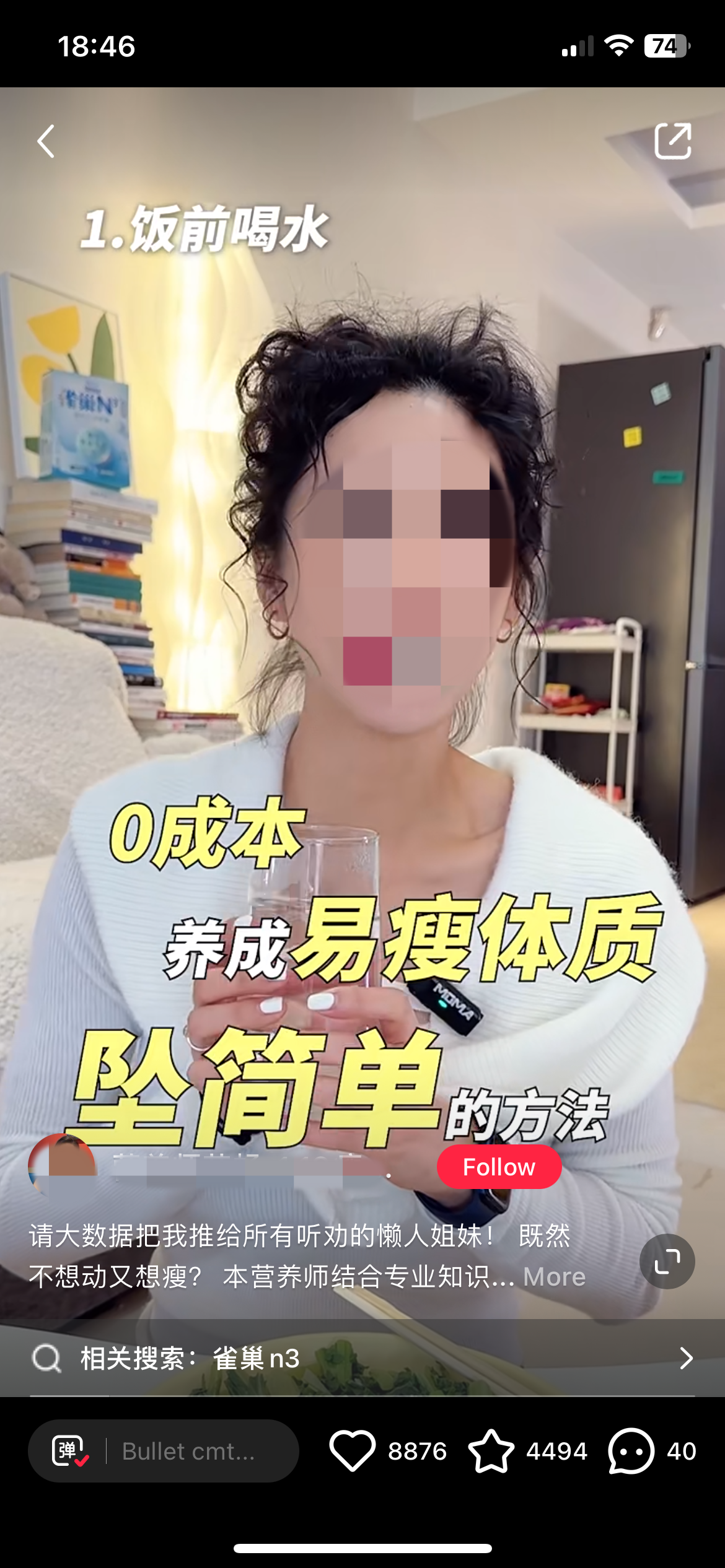} 
        \caption{}
        \label{fig:video}
    \end{subfigure}
    \begin{subfigure}{0.2\textwidth}
        \includegraphics[width=\linewidth]{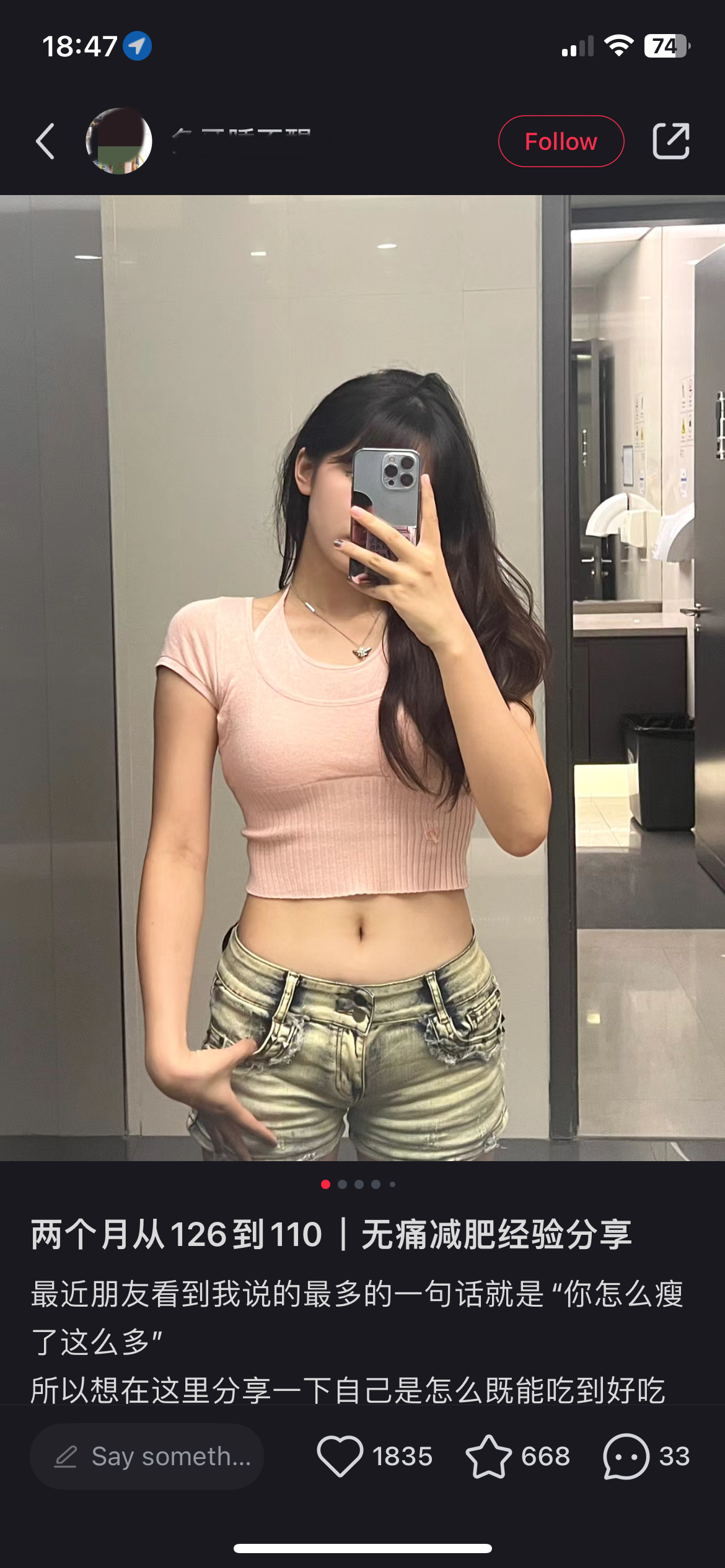} 
        \caption{}
        \label{fig:photo}
    \end{subfigure}
    \begin{subfigure}{0.2\textwidth}
        \includegraphics[width=\linewidth]{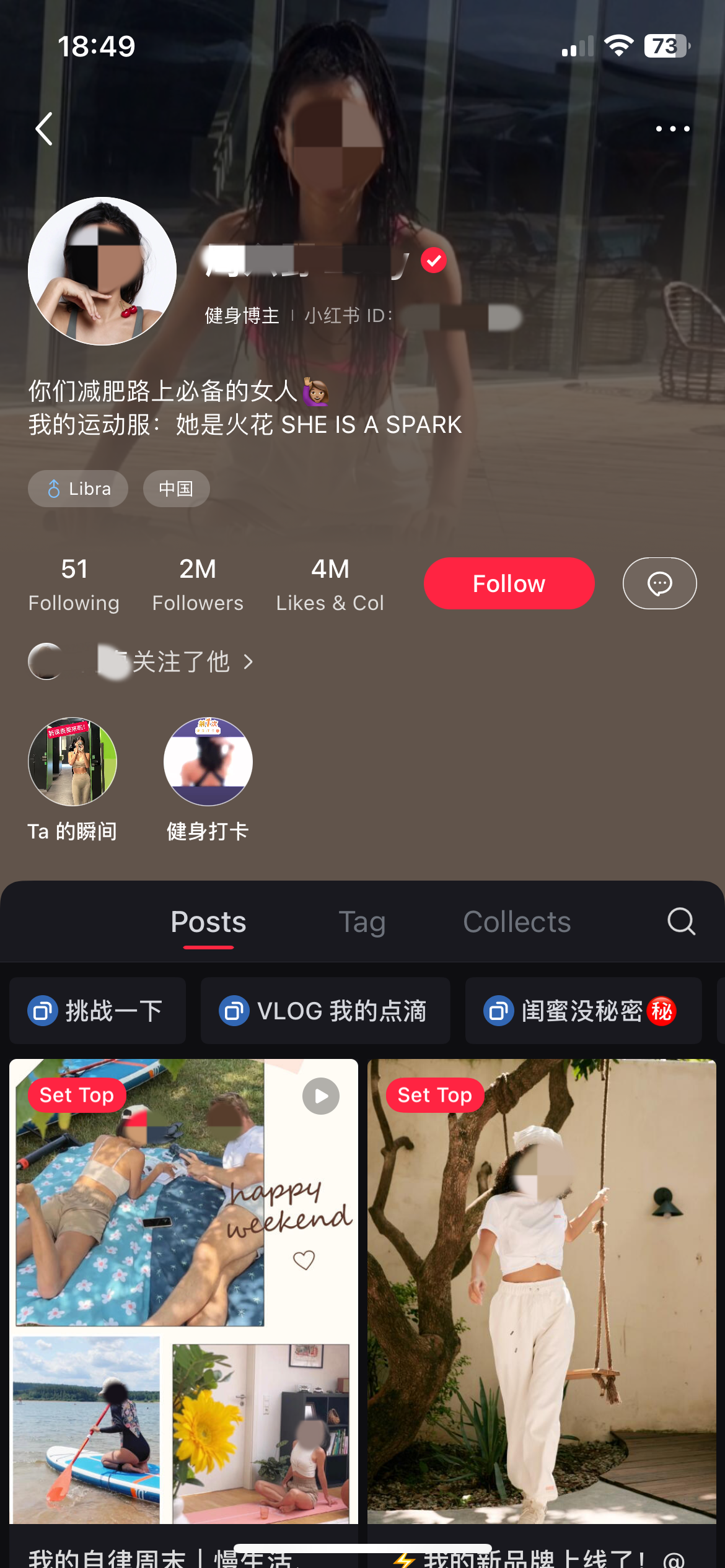} 
        \caption{}
        \label{fig:profile}
    \end{subfigure}
\caption{RedNote Interfaces: (a) \textit{Body management search results page}: Users search for terms like ``weight loss,'' and posts show titles such as ``Lazy people can easily gain weight without exercise'' or ``Weight loss tips and tricks,'' along with views and likes. (b) \textit{Video post}: A creator shares a ``zero-cost'' method to develop a ``slim body type,'' with captions summarizing the advice. (c) \textit{Photo post}: A user shows before-and-after photos documenting a two-month weight loss journey, sharing tips in the caption. (d) \textit{User's profile}: Displays follower count, total likes, bio, and a collection of posted photos and videos.}
\Description{These figures show the main features of Little RedNote Book}
\label{figure1}
\end{figure}

RedNote has emerged as both a social and e-commerce platform \cite{liu2022discipline}. With over 90 percent of users on the platform identified as female, it is also referred to as an online `SHE' community \cite{chi2022users, zhang2021use}. Established in 2013, RedNote began as a platform for users to share their purchasing experiences, later transforming into a comprehensive lifestyle-sharing hub covering an array of interests, including fashion, travel, wellness, and weight loss \cite{liu2022discipline}. The keyword search function (shown in Fig. \ref{fig:search}) on RedNote allows users to input specific terms related to their interests. Users can then filter results based on relevance, tailoring their searches to find the most current and pertinent information, which comes with different formats, such as text-based posts (Fig. \ref{fig:photo}) with images, videos (Fig. \ref{fig:video}), and live streaming \cite{wang2022impact}. On RedNote, some users have gradually evolved into influencers with a large number of followers. Compared with posts by normal users, those by influencers involve more high-quality images and engaging videos, capturing the attention of viewers. Influencers often collaborate with brands by integrating product recommendations into their content \cite{niu2023study}. Figure~\ref{figure1} illustrates the interface of the application, showcasing the dynamic nature of user-generated content. Users on RedNote engage with each other through comments, private messages, and likes, fostering a vibrant and interactive community.

Because RedNote has positioned itself not only as a social media but also as an e-commerce platform, most previous work has predominantly studied the commercial aspects, marketing models, and female consumerism on it \cite{ren2021precision, gong2022research, lian2021perspective}. This study instead explores how RedNote, positioned as an information-sharing platform, impacts users' decision-making. We focus on body management and related, highly searched topics such as weight loss, aiming to understand how such a female-dominant platform influences users' information-seeking and evaluation.

\subsection{Online Body Management Communities}
In HCI, the Online Health Community (OHC) has gained much attention. Individuals increasingly turned to online communities for information and support \cite{ballantine2011help, hwang2010social}, which can provide them with access to a broader social network and diverse perspectives \cite{tanis2008health, wright2003health}. Health information seeking in HCI covers a wide range of topics from pregnancy \cite{gui2017investigating}, obesity \cite{mejova2018information}, diabetes \cite{zhang2017supporting}, to childcare concerns \cite{kaur2020understanding}, etc. Body management behaviors such as tracking diet and exercise have been well-studied \cite{thomaz2015inferring}, with studies exploring the technology designs for effective monitoring and tracking of progress. Moreover, the adoption of digital tools for food intake documentation, called digital journeying, is a proven strategy in weight loss. For example, researchers studied how the MyFitnessPal app could facilitate weight loss by helping people document food intake \cite{gordon2019goal}.

Other than tracking, information seeking is another important research area in HCI. Users tend to use online health forums for information and support. For example, \cite{li2014losing} investigated users' participation experience in a weight loss community called DropPounds. Chancellor and his colleagues \cite{chancellor2018norms} compared social support's impact on users' behavior change in two weight loss communities on Reddit, pointing out that context and community norms, not just word choice, determine the meaning and impact of online support. Hwang et al. \cite{hwang2013website} identified a positive correlation between weight loss and participants who logged their weight at least four times. Moreover, health professionals are also engaged on online platforms to provide information. For example, \cite{may2017weight, munson2013sociotechnical} studied how people used Twitter to interact with peers or health professionals for health information and support. 

While extensive research has examined how users seek and assess online health information, most existing work has focused on health websites or text-based forums and explored general criteria such as accuracy, credibility, and usefulness \cite{cunha2016effect, hwang2013website}. However, less is known about how information evaluation (the process through which users judge the credibility, relevance, and personal applicability of information) unfolds on socially and commercially mediated platforms like RedNote, where influencer culture, commercialization, and peer interactions intertwine. In particular, little research has examined how different information modalities (e.g., video vs. text) shape users’ information-seeking and evaluation behaviors in this context. To address these gaps, our study investigates how video and text-based posts influence users’ engagement with body management information and how the platform’s commercial atmosphere shapes their attitudes toward social media content.

\subsection{Challenges Users Encountered in Seeking and Evaluating Health Information}

Users encounter various challenges when seeking health information online. Studies assessing the quality of websites related to obesity revealed that the information presented was frequently inaccurate, misleading, or conflicting. For instance, research by Laz et al. \cite{laz2011association} indicates that adolescents obtaining weight loss information online are more likely to engage in unhealthy weight loss behaviors compared to non-seekers. Some interactive online forums offer health information that is either misleading or potentially harmful \cite{hwang2007quality}. Despite the engaging and visually appealing nature of user-generated content on forums or blogs, such content is often authored by individuals lacking medical or health professional experience and operates without regulation \cite{cline2001consumer, culver1997medical, moorman2020use}. Consequently, it frequently contains messages inconsistent with standard medical practices and recommendations \cite{cline2001consumer, culver1997medical, moorman2020use}. For example, an analysis of content from nutritional websites uncovered that platforms advocating a `healthy lifestyle' and `inspirational fitness' often support unhealthy eating behaviors and attitudes, such as dieting, practicing dietary restraint, conveying guilt-inducing messages, and engaging in problematic eating \cite{boepple2014content, moorman2020use}. Users also face challenges in evaluating information. 
Although challenges in health information-seeking have been explored in previous studies, none have specifically addressed the unique obstacles faced by users on a female-dominant platform where beauty, fashion, and influencer marketing significantly shape the user experience \cite{shi2021role, zhang2022influencer, wei2023re}. To fill this gap, the present study investigates the distinct challenges that RedNote poses for users engaged in body management practices.

\section{Method}

\subsection{Recruitment Procedure}
Two researchers posted recruitment information on RedNote using their personal accounts. The recruitment information included the basic information of the researchers, the purpose of the study, the format and estimated length of the interview, and the compensation information. Interested RedNote users either directly messaged the researchers or left comments under the post, after which they were contacted by the researchers. The researchers asked three screening questions before scheduling interviews: 1) if they are equal to or older than 18 years old, 2) if they have experience seeking body management information on RedNote, and 3) whether they identify themselves as female. Participants who answered all questions with ``yes'' met the study requirements and received emails from the two researchers to schedule online interviews. Eventually, a total of 18 participants completed the interviews. They had been using RedNote for periods ranging from five months to six years. All participants were identified as domestic Chinese, born and raised in mainland China, although some had migrated overseas for study or work at the time of the interview. Table~\ref{tab:demo} shows the demographic information of the participants.

\begin{table*}[ht]
\caption{Participants demographic information}
\label{tab:demo}
\resizebox{\textwidth}{!}{%
\begin{tabular}{llll|llll}
\toprule
\multicolumn{1}{c}{Participant ID} &
  \multicolumn{1}{c}{Gender} &
  \multicolumn{1}{c}{Current Residency} &
  \multicolumn{1}{c|}{App Usage} &
  \multicolumn{1}{c}{Participant ID} &
  \multicolumn{1}{c}{Gender} &
  \multicolumn{1}{c}{Current Residency} &
  \multicolumn{1}{c}{App Usage} \\ \midrule
P1 & Female & US      & 5 months  & P10 & Female & US          & 4 years \\
P2 & Female & US      & 4 years   & P11 & Female & PRC         & 3 years \\
P3 & Female & PRC     & 3 years   & P12 & Female & PRC         & 5 years \\
P4 & Female & UK      & 3 years   & P13 & Female & UK          & 4 years \\
P5 & Female & PRC     & 3 years   & P14 & Female & PRC         & 4 years \\
P6 & Female & PRC     & 2 years   & P15 & Female & New Zealand & 1 year  \\
P7 & Female & Germany & 3.5 years & P16 & Female & UK          & 1 year  \\
P8 & Female & PRC     & 4 years   & P17 & Female & PRC         & 6 years \\
P9 & Female & Ireland & 6 years   & P18 & Female & PRC         & 2 years \\ \bottomrule
\end{tabular}
}
\end{table*}

\subsection{Interview Design and Analysis}

The interview questions were iteratively developed by researchers based on the relevant literature and two pilot interview results. The questions were developed in English and translated into Chinese for the formal interview. The study used a semi-structured interview format, where additional questions were generated based on responses to the main questions, and participants were allowed to discuss other tangential topics. The interview questions covered topics related to the search for information on body management, as well as comparing and evaluating RedNote content. Example questions include ``\emph{How do you feel about the body management topics on this platform?}'' ``\textit{Could you tell me the challenges or issues you encountered when seeking body management information?}'' and ``\textit{How do you decide whether you would trust the information or not?}''

We conducted an in-depth qualitative analysis using an inductive thematic analysis approach. In the initial phase, three researchers independently and simultaneously coded six interview transcripts to establish a shared interpretive baseline and to surface diverse perspectives. This open coding stage produced approximately 150 distinct codes. The codes were then collated in an Excel spreadsheet and iteratively organized into fifteen preliminary themes. Each researcher subsequently performed an in-depth review of a subset of the codes (around fifty per researcher), identifying representative quotations that best illustrated the meanings of each theme. For example, codes such as ``video illustrates actions clearly and increases trust,'' ``text and images easier to review and recall,'' and ``streaming perceived as inflexible'' were grouped under the theme ``Different Content Modalities Influence Information Evaluation.''

To enhance analytic rigor and reliability, the three researchers met regularly to compare interpretations, discuss discrepancies, and refine the code definitions and thematic boundaries. Disagreements were resolved through discussion until consensus was reached. A senior researcher supervised this process, reviewing the evolving codebook and theme structure to ensure that the analysis accurately reflected the data, that relationships among codes were systematically examined, and that the final themes were coherent and well substantiated by evidence.

\section{Findings}

\begin{table*}[ht]
\centering
\caption{Summary of Themes, Subthemes, Representative Codes, and Number of Participants (n: 18)}
\resizebox{\textwidth}{!}{%
\begin{tabular}{p{0.25\textwidth} p{0.25\textwidth} p{0.42\textwidth} p{0.04\textwidth}}
\toprule
\textbf{Theme} & \textbf{Subtheme} & \textbf{Representative Codes (condensed)} & \textbf{n}\\
\midrule

\textbf{Trust in Influencers vs. Everyday Users}
& Commercial activities can decrease users' trust
& Perceived over-commercialization; authenticity over popularity; skepticism toward monetized posts
& 8 \\[3pt]

& Professional certificates can increase trust
& Value of credentials; preference for professional trainers; scientific framing increases credibility
& 6 \\[3pt]

\midrule

\textbf{Different Content Modalities Influence Information Evaluation}
& 
& Videos aid understanding; text easier to revisit; dislike AI voice or over-edited visuals; visual appeal affects trust
& 12 \\[3pt]

\midrule

\textbf{Users’ Concerns and Challenges Regarding Body Management on RedNote}
& Unhealthy/Extreme Measures and Struggles to Discern Reliable Methods
& Awareness of extreme diets; difficulty judging scientific validity; preference for moderate approaches
& 8 \\[3pt]

& Low Commitment for Social Media Instructions
& Saved but unpracticed tips; lack of supervision; low long-term adherence
& 6 \\[3pt]

& Body Anxiety Induced by RedNote
& Gendered comparison; male misunderstanding; appearance pressure and social anxiety
& 10 \\
\bottomrule
\end{tabular}%
}
\label{tab:code}
\end{table*}

The themes and subthemes we found are summarized in Table~\ref{tab:code} As an e-commerce platform, RedNote is a platform where commercialization blends with social engagement. Users can create posts, interact with others through comments, send private messages, start group chats, or do live streaming, etc. At the same time, RedNote provides a space for people to promote and sell products, especially for those who have a good number of followers. For example, some influencers on RedNote are not only social media users who post and interact with others, but also product advertisers who use their influence to advertise products. Influencers on RedNote, especially those who have a large number of followers, often collaborate with multichannel networks (MCNs) to expand their reach, streamline brand partnerships, and ultimately enhance their monetization potential \cite{liang2024manufacturing}. MCNs refer to these agencies that represent groups of social media influencers, connect influencers with brands, and provide support in content creation, campaign management, and audience growth strategies. In exchange for these services, influencers are typically required to share a portion of their revenue with MCNs \cite{zhang2024contesting}. Due to the business model, the presence of influencers who promote products related to body management (e.g., diet products, fitness equipment) affects the trust, engagement, and authenticity of information. Users have to learn how to navigate the blurred line between genuine advice and product promotion, a challenge unique to RedNote's ecosystem.

\subsection{Trust in influencers versus everyday users.}
On RedNote, influencers owning a large number of followers have great power to disperse messages as they are often considered more reliable resources than normal users. Our interviews confirmed that when it comes to evaluating information, popularity could mean usefulness, yet it is not always the case.

\begin{quote}
    I judge the credibility of this information by looking at its number of followers, mainly the number of fans. If an account has a particularly large number of followers, I think it is a relatively trustworthy account. -P6
\end{quote}


As P6 described, a large number of followers can be a strong endorsement that a blogger is trustworthy. P10 also linked popularity with professional content, as these certified bloggers would make efforts to maintain their fans.
\begin{quote}
    Firstly, if the source of the information comes from a very professional, certified blogger on the platform specializing in body management, I would tend to trust them more. I think they would be more professional, and the authenticity of the content they publish would be stronger. Given their large fan base, they would not likely jeopardize their own reputation. -P10
\end{quote}
This connection between popularity and perceived authority of a blogger is widely observed on social media platforms, and body management posts are no exception. However, this popularity-centric dynamic is not without its counterpoints. Some participants have displayed a preference for content that is more `down-to-earth', questioning the fact that content produced by influencers is heavily or even overly edited.
\begin{quote}
    I would not choose to follow particularly big bloggers on RedNote. Smaller bloggers seem more lively, more real, and more alive. It feels like they give you the sense that they are genuine people. Big bloggers, on the other hand, might feel like they produce content in a conveyor belt fashion. Can I say that? -P17
\end{quote}
As P17 complained, the content of popular bloggers can appear manufactured and unreal. Instead, non-influencers who are everyday users may gain more trust. Their posts are typically less polished and more reflective of everyday life. P18 gave a vivid example of contrasting posts from different bloggers, depreciating posts that overly highlight visual appeal:
\begin{quote}
    I prefer posts that are particularly authentic. For instance, if you are documenting your fitness and body management journey, I don't really like posts that focus too much on tight clothing or heavily edited body photos. I favor those that are genuinely casual, such as a quick snap of the weighing scale, a picture of a meal, or sharing honest thoughts. Or maybe a close-up showing something like sweat from a workout. Those are the kinds of thing that I tend to like. But if your post first shows a highly edited selfie of your face or profile, I generally will not click on it. -P18
\end{quote}

`Down-to-earth' posts may resonate better with users seeking authenticity and relatability. This preference suggests some users' awareness of the manipulation exerted through popularity. Instead, they desire content that mirrors the real-life experiences of ordinary individuals:
\begin{quote}
    I am willing to believe in authentic photo and text content. For example, if someone shares their personal journey and thoughts in a format such as a memo, that is the most convincing to me. And then, secondly, things like photo check-ins, showing what they ate, I also find those to be more realistic. -P15
\end{quote}
The dichotomy between influencer-driven content and layman posts highlights a complex interplay in user preferences. While influencers can leverage their social capital to increase views and comments, there may be a growing need for content that is unfiltered and genuine. This trend underscores that the credibility of body management can come from both popularity and authenticity.

\subsubsection{Commercial activities can decrease users' trust}


Many of us experience a sense of intrusion when a salesperson unexpectedly knocks on our door, interrupting our daily routines. It seems that this sentiment extends to social media platforms. Users seeking body management advice are skeptical of bloggers who are overtly linked to commercial endeavors, such as advertising.
\begin{quote}
    When I visit a blogger's page, if they are a fashion blogger with hundreds of thousands of followers, I tend to think that they might be promoting a company's products. So I check if the products they recommend come from the same company. If it is always the same company, I might not be very inclined to trust the blogger. -P12
\end{quote}
This antipathy towards commercial activities also contributes to users' preferences for less popular bloggers, as influencers are often targets of advertisement. As P10 explained, these bloggers are manipulated by monetary incentives and therefore may not post trustworthy body management information. This association can significantly diminish the perceived credibility of their content. 
\begin{quote}
    If I see that they post about body management, but their other posts are mainly about recommending diet replacement meals or promoting products, giving off a commercial advertising vibe, I probably will not trust them much. I am concerned that your opinions might be influenced by your own financial interests. -P10
\end{quote}
\subsubsection{Professional certificates can increase trust}
Clearly, popularity alone is insufficient to determine the trust of users in body management posts. In RedNote, users could prove that they have professional certificates, licenses, or other credentials. They can be applied to be verified as professional accounts. Once verified, the platform labels these accounts as professional, helping users distinguish expert content from regular user-generated posts, enhancing the credibility of the information shared, particularly in areas like health and body management. Among the diverse range of influencers, those who present professional content can often overcome the counterpoints of popularity and emerge as particularly credible sources.
\begin{quote}
    I follow bloggers who are certified fitness coaches. Although I am not sure about the actual value of these certificates, I think their training methods are more scientifically credible. I also pay attention to bloggers who guide you on exercise types, or those whose content is very clear, like how to slim down thighs, work on the waistline, or shape specific body parts to improve your physique. -P4
\end{quote}
As P4 pointed out, certified trainers tend to produce more scientific and reliable body management content. Their posts are often characterized by well-organized and clearly instructed body management tutorials. P4 gave another example of professional bloggers who endorsed their body management posts with research literature and expert opinions:

\begin{quote}
    I now tend to prefer bloggers like Xixi Chen, who sometimes reference scientific literature or consult professionals. Although I don't completely trust these professionals when they provide evidence-based information, I feel it is more reliable, and indeed she does this. -P17
\end{quote}

By producing professional and high-quality body management content, these bloggers provide a layer of authority and reliability that separates them from other influencers. This phenomenon highlights the importance of professionalism even in body management posts on social media.

\subsection{Different content modalities influence information evaluation}

The features of the RedNote platform play an important role in how users connect with the body management content and perceive its effectiveness. By focusing on visual storytelling with photos and short videos, the platform allows users to share clear before-and-after contrasts that resonate with people looking for inspiration for body image. This simple and visually appealing approach makes it easy for users to see how specific body management methods can work, making the content feel relatable and motivating. Throughout our interviews, participants shared that they were especially drawn to posts showing before- and after- appearance pictures, as these comparisons offered a real, tangible sense of how effective the body management strategies they had used.

\begin{quote}
    My preference is usually for those with more obvious before and after comparison photos, accompanied by content on daily fitness routines, and then, if the fitness journey or body management period is quite long, this tends to feel more authentic. -P16
\end{quote}

In addition to posting pictures, videos provide a visual and auditory experience that excels at demonstrating physical exercises, making it easier for viewers to understand and replicate movements accurately. This is particularly crucial in exercise routines, where proper form and technique are essential not only for effectiveness but also for safety. This format allows for a more comprehensive understanding of exercises, including nuances in posture and tempo, which are difficult to convey through text or static images alone.

\begin{quote}
    I think the videos are definitely more detailed. For example, they teach you a specific training movement, including the position of your feet, the spacing between your legs, and which muscle groups to engage. They point these out very specifically, unlike text and images, which can be quite cold and vague, merely stating which muscle groups to use without specifying which ones. -P1
\end{quote}

However, the importance of pictures and text in body management posts cannot be understated. Graphic and textual posts possess the unique ability to provide concise snapshots and summaries of complex information, making them invaluable for quick reference and clarity.  
\begin{quote}
    I would prefer if bloggers, who attract me through their videos, could also make a graphic and text summary of what they've talked about. A graphic and text summary would make it easier for me to follow. With videos, sometimes I have to scrub through the video to re-create the information. But with graphics and text, I can just open it up and see exactly what I am looking for. -P15
\end{quote}

For users seeking to understand or follow a workout routine, these formats offer a straightforward and accessible means of accessing vital information, while videos serve more as an entertaining method.
\begin{quote}
    The graphic and text content is mainly about the fitness routines, such as which exercises they did. With graphics and text, there is a clear outline of what to do on Monday, Tuesday, etc. As for videos, I watch them while eating or when I am idle to pass the time, and they focus mainly on dieting. -P14
\end{quote}

As mentioned in P14, pictures and text can effectively illustrate exercises or present body management plans with visual clarity, which is particularly beneficial for users who require a quick overview or a handy guide that they can refer to anytime and anywhere. It should be noted that participants have expressed a dislike for videos that use AI narrators. This aversion is largely due to the lack of personal touch and authenticity that AI voices can sometimes convey.

\begin{quote}
    If the blogger does a voiceover, I have higher expectations for their tone and intonation. The type of voice modulation that's computer-generated is very popular on RedNote, featuring the Mandarin dubbing on TVB (Television Broadcasts Limited, Hong Kong). However, I prefer not this kind of dubbing, but rather something that's authentic and real. -P16
\end{quote}

\subsection{Users' Concerns and Challenges Regarding Body Management on RedNote}
\subsubsection{Unhealthy/Extreme Measures and Struggles to Discern Reliable Methods}
The prevalence of unhealthy and extreme body management measures in RedNote has been a major concern among participants. Along with vast amounts of health and fitness content, there is a significant amount of advice that is not scientifically supported, potentially leading to unhealthy practices.  For example, P3 talked about her frustration with the nonscientific nature of some body management posts on RedNote: 

\begin{quote}
   They emphasize that one must eat healthily and cleanly, but they also say that you cannot do this, or that there are so many things that you need to avoid. It goes clearly against the concept of a scientific diet. I find myself disliking these things. There is so much that you supposedly cannot do or eat, at that rate, one might as well not live. -P3
\end{quote}

Although the participant's words may be exaggerated, she described a phenomenon where posters claimed their measures were healthy, but what they advocated were actually unhealthy and overrestrictive, which she believed were self-contradictory and lacked a scientific foundation. Such methods of body management may not only severely affect quality of life, but also have certain negative impacts on health. As the impact of such information without scientific evidence can be detrimental, participants reported instances of being misled by popular but unverified body management methods. Specifically, the promotion of certain diets or fasting routines without proper nutritional guidance poses the risk of health complications and nutritional deficiencies. 

P16 discussed how, despite her personal belief that the dieting and fasting posts on the platform were unhealthy, she was still influenced by them and ended up eating less in her daily routine. This reflects the widespread influence of this content, even among those aware of its potential harms. Similarly, P9 expressed concern for younger users of RedNote, warning against the prevalent body management methods that often encourage eating very little. ``\emph{Although they might be effective, truly healthy body management methods are rare, and most are quite unhealthy. I don't think it is good, especially for younger individuals who might still be in their adolescence. It is a very misleading guide for them.}'' P18 mentioned the difficulty of finding critical information about the measures when searching on the platform. 


\begin{quote}
    For example, you will not find why the `Copenhagen diet' might be unhealthy. You only see this popular method itself, most instructing you on how to do it. The best you can get is some more scientific modified versions based on the original or the effects, but you don't see information about its downsides like muscle loss. -P18
\end{quote}

The participant emphasized that while detailed methodologies of measures are readily available on the platform, critical information about potential adverse effects is absent. This lack of comprehensive insight could lead to a noticeable decline in one's physical well-being, especially if combined with a decrease in physical activity.

While it's crucial to avoid misleading information, participants also highlighted the challenge in discerning scientifically backed, reliable body management information from the multitude of content available. The vast and varied content of the platform makes it difficult for users to identify which is trustworthy and what is not. As P9 mentioned, ``\emph{There's an overwhelming amount of information, and there are so many fitness and body management methods that it can be difficult to discern which one is scientific, which one is truly effective, and ultimately, which regimen to follow.}''


\subsubsection{Low Commitment for Social Media Instructions}
Despite the initial enthusiasm for RedNote body management advice, many users find it difficult to maintain a consistent routine. When asked about the effectiveness of the body management methods learned from the platform, many participants reported not seeing any significant changes, often attributing this to lack of regular practice, typically only engaging in the activity once or twice. As P11 noted, ``\emph{It's hard to say if there's any effect because I don't follow a regular exercise regimen. So, it is difficult to judge any results.}'' 

Regarding the lack of follow-up, some participants explained that it could be due to the absence of supervision and the casual approach to exercises found in RedNote. P13 compared her experiences of following RedNote methods with attending a gym with trainers. 

\begin{quote}
    I have not found a long-lasting body management method on the platform, but this might not be related to the content itself. It is often due to personal reasons, such as being lazy and skipping a few days of exercise. However, I don't feel this way with a gym trainer, as they insist that I complete the exercises. With RedNote methods, I usually practice them at home by myself, and I often find them too tiring. -P13
\end{quote}

Compared to gyms, RedNote's methods lack the structured environment and accountability provided by trainers that are more conducive to sustained effort. Furthermore, some participants pointed out that while searching for various body management information on RedNote, it is more about looking up exercises sporadically rather than following a systematic body management plan. P11 suggested, ``\emph{I exercise when I feel like it, but I don't strictly follow any particular regimen or have a systematic approach to body management.}'' This implies a more casual approach to exercise, lacking the structure typically associated with more successful body management strategies.

In addition, many participants explained that the ease of saving posts with the intention to return but not doing so due to natural laziness. P12 expressed, ``\emph{I always save the posts, but none are consistently followed through; they just collect dust in my collection. People have an inherent laziness and that often gets in the way.}'' This statement reflects a common issue where the intention to engage with saved content does not translate into action, primarily due to procrastination or lack of motivation.

\subsubsection{Body Anxiety Induced by RedNote}
Content on RedNote can also contribute to body anxiety. The platform, with its vast amount of body management and fitness posts, can both amplify existing insecurities and create new ones, particularly through the promotion of ideal body types or through comparisons with others. The accessibility to such content often exacerbates users' concerns about their body shapes, leading to increased anxiety and pressure to conform to certain standards. P13 noted cultural differences in body perception, stating that she felt less body anxiety when living abroad compared to being in China. 

\begin{quote}
    When I was studying abroad, I felt good about how I looked and thought I looked good in what I wore, and others seemed to think so too. However, after returning to China and looking at those photos, I realized that I was actually so fat and not as pretty as I thought. My friends also mentioned that the topics we discussed changed significantly after we returned to China. In China, there seems to be more emphasis on dieting or how some girls look great and are popular. -P13
\end{quote}

The participant suggested that the cultural influence and the prevalent standards of beauty in China were attributed to the change in perception of her body. When the researchers asked whether she believed RedNote plays a significant role in creating this environment, the participant suggested that while the platform itself might not be the direct cause of anxiety, it certainly acts as a catalyst in the culture already focused on slim shapes and fine appearances. 

P7 discussed societal and platform-induced anxiety more in-depth, particularly emphasizing the unrealistic body weight standards perpetuated in RedNote. 

\begin{quote}
    I think body anxiety is primarily due to social pressures and RedNote especially exacerbates this anxiety because there are so many beautiful girls on the platform. There's also a plethora of body management experiences shared, detailing how individuals slimmed down significantly... This makes me anxious, feeling like all the girls online are beautiful and slim, and it gives me the impression that I, too, could be prettier if I put in more effort. -P7
    
\end{quote}


P11 remarked on how some posts on RedNote are explicitly designed to induce body anxiety, with targeted messages that play on users' insecurities about their bodies. ``\emph{On RedNote, there are posts specifically designed to create body anxiety, such as `Summer is here, do you really want to wear your crop tops with a belly?'} The participant expressed discomfort with these tactics, stating, ``\emph{These kinds of headlines are created for traffic. They make me very uncomfortable. I believe they are deliberately manufacturing body anxiety for women, prompting you to follow their posts for some exercise training or to buy their body management products.}'' According to her, this manipulation of users' insecurities for engagement or motivation is a notable concern that contributes to the overall environment of body anxiety on the platform. 

Even when posts are not intended to cause anxiety, the pervasive culture of comparison and idealization in RedNote often results in significant body satisfaction. For example, P15 spoke about the platform algorithm and how, despite attempts to filter content, the abundance of beauty and body management posts still leads to body anxiety. ``\emph{Just seeing these beautiful and slim girls on the platform can induce anxiety about my own body and looks.}'' The constant exposure to idealized bodies and lifestyles creates a relentless pressure to conform and improve, sometimes at the expense of one's mental health. 

 
Lastly, some participants emphasized the misleading influence of models who unintentionally promote unrealistic body perceptions for ordinary people on the platform. P16 mentioned, 
\begin{quote}
    Some models post about how they maintain an extremely low weight, including what they eat for every meal. I think this can be somewhat misleading. Because models have job requirements to maintain a certain figure and weight, this does not necessarily apply to the general population. -P16
\end{quote} 
The participant suggests that most users only see their beauty, not understanding that different professions and individuals have different needs, and what they share might not be suitable for the majority. 

\section{Discussion}



\subsection{The Mixed Social Commerce and Online Community of RedNote Impact Users' Decision}
\subsubsection{Influencer culture} RedNote serves a dual purpose as both a life-sharing and e-commerce platform. This unique identity positions it not only as a hub for seeking and sharing information but also as a business-centric platform. With the majority of users being women, there is a notable increase in social media influencers, predominantly young women in beauty and fashion \cite{abidin2016agentic, duffy2019gendered, senft2008camgirls}. Individuals with a significant social media following can monetize their online posts through brand collaborations and sponsored content \cite{niu2023study}. However, rather than adopting an overt promotional approach, RedNote influencers often employ storytelling strategies to engage users and influence their purchasing decisions \cite{wei2023re}, blurring the lines between personal expression and commercial promotion. Although this content marketing strategy proves to be effective for businesses, it presents challenges to users when discerning between advertisements and genuine information sharing. This complexity increases the difficulty for users to assess the reliability of the information shared in RedNote. Throughout our interviews, it was evident that users actively sought to avoid advertisements and took measures to move away from business-driven content aimed at promoting body management products. 

\subsubsection{Popularity VS. Down-to-earth} Furthermore, previous work indicated that online celebrities with a larger following are considered more reliable in information sharing compared to those with fewer followers \cite{das2017celebrities}. Likewise, our interviews revealed that certain users tend to trust advice on body management when it comes from Internet microcelebrities with a significant following. This trust is often linked to the perceived high quality of these celebrities' posts, which showcased more aesthetically pleasing images \cite{djafarova2019instafamous}. However, some participants expressed a preference for lesser-known content creators, valuing the authenticity of their posts without heavily edited pictures because this type of content is perceived to be more genuine life experience sharing rather than covert advertisements. The contrast in user preferences suggests that although a substantial following may enhance trust, there exists a considerable user segment that prioritizes authenticity over popularity, likely due to the unique mixture of life-sharing and e-commerce content on RedNote. Hence, when developing features or policies related to content creation and promotion, platforms should take into account these diverse preferences. 

\subsubsection{Information provider's authority} Moreover, our findings showed that no matter whether users lean toward posts from popular influencers or those from more down-to-earth creators, all place importance on the professionalism of content creators. Once they can verify that the posts come from scientific sources or bloggers with fitness certifications, they are more likely to follow the suggestions. As a result, we suggest that the RedNote platform enhance the user experience by helping them differentiate between professional and non-professional content creators. For example, RedNote could adopt a labeling system \cite{spradling2021protection} that clearly identifies content creators with health-related certifications or scientific backgrounds, and the labeling system can also label content with scientific evidence as "professional answer". Such targeted labeling would provide users with a more precise and transparent means of assessing the credibility and expertise of the individuals they choose to follow, thereby promoting informed decision-making and trust within the platform's content ecosystem.

\subsection{Media Modalities and Content Interact to Shape Users' Information Evaluation}
Media and content dynamically interact to shape user viewing experiences. Previous research suggests that both videos and texts possess unique strengths in conveying information. Visual elements, such as videos and images, effectively engage viewers, evoking a stronger emotional response \cite{yadav2011if}. In contrast, text excels in conveying information in more structured and informative presentations. Multimodal information sharing in RedNote enables users to interact with content in various formats according to their preferences. Our findings indicate that RedNote users prefer to watch videos when seeking exercise instructions, as videos provide clearer demonstrations of fitness movements. However, when users are looking for information about dieting, such as a food plan, they prefer reading text accompanied by pictures for ease of comprehension and reference. Interestingly, even users who prefer obtaining information through videos express a desire for text summaries to help them condense key information. Given these insights, a potential recommendation is for the RedNote platform to implement artificial intelligence techniques to generate abstractive text summaries from videos \cite{atri2021see}. This would empower users to generate concise summaries based on the videos they've viewed, enhancing their ability to process and retain information. 

Previous work showed that perceived authenticity increases feelings of liking \cite{nah2022appeal}. In alignment with this finding, our study confirmed that authenticity stood out as a key factor impacting users' preferences for certain videos on platforms. A noticeable shift on the RedNote platform involves an increasing number of videos featuring AI voice-overs instead of real human voices, which has sparked resistance among users. To address this, we propose that platforms, RedNote included, consider implementing a recommendation system. This system would personalize video suggestions based on users' past preferences, focusing on the intrinsic features that resonate with them. Moreover, to improve overall authenticity, platforms could manage the prevalence of AI-generated videos by reducing traffic support, thus decreasing the popularity of such content.

\subsection{Physical and Mental Labor Required as Users Seek Information}

Most previous work in HCI has examined weight loss as a clinically guided process to address medical conditions such as obesity or disordered eating \cite{ludwig_recommending_2021, eikey_its_2017}. In contrast, our study explored how body management has become a lifestyle practice—particularly among women—motivated by goals such as improving appearance, enhancing well-being, and maintaining a desirable body shape. Compared with medically supervised programs that provide professional guidance and structured routines \cite{freyne_mobile_2010, ryan_understanding_2022}, lifestyle-oriented body management requires individuals to take on substantial personal responsibility for seeking, evaluating, and implementing information found online. Our findings reveal that engaging with body management information on RedNote entails both \textit{physical} and \textit{mental} labor. Physically, participants invested considerable embodied effort in translating online advice into daily practices—experimenting with recommended workouts, adjusting diets, and continuously monitoring bodily changes to assess feasibility and effectiveness. These ongoing routines demanded persistence, self-discipline, and time, transforming digital engagement into sustained forms of bodily work. Mentally, participants described the process as psychologically taxing. The platform’s e-commerce orientation and influencer culture amplify appearance-related expectations and comparison. Many participants reported experiencing heightened body anxiety as they navigated a constant flow of promotional and user-generated posts emphasizing thinness and rapid transformation. Exposure to unrealistic or extreme body management methods further intensified these feelings, leaving users oscillating between motivation and self-doubt. Together, these forms of embodied and mental labor illustrate how body management on RedNote extends beyond information seeking or consumption. It becomes an ongoing cycle of evaluating, performing, and emotionally negotiating one’s body, shaped by the platform’s commercial atmosphere and social interactions.

\subsection{Design implications}
\subsubsection{Organization of body management posts}
RedNote benefits from the abundance of user-generated content. Although various posts can satisfy various user needs, they can also create clutter and information overload. Users may find it difficult to sort through numerous reviews or comments to make informed decisions, especially for body management posts. Our findings suggest that the user experience in information search could be improved by a more customized organization of results. Specifically, RedNote users paid significant attention to nuanced characteristics such as media format, fan size of the blogger, previous posts of the blogger, and even the gender of the blogger. Therefore, the social media search interface may be designed to enable users to customize the criteria for sorting and filtering the results of a search query. 

For body management posts specifically, it is also necessary to determine if the information provided is scientific or healthy. However, evaluating whether a body management strategy is scientific is beyond the ability of most users, especially considering the uneven quality of user-generated content. Therefore, the design of RedNote can be improved by adding post labels that highlight professional or good-quality posts. These labels can be assigned using expert evaluation or interaction metrics, such as user votes. Through these labels, the platform could encourage bloggers to post high-quality content that promotes healthy body management strategies.

\subsubsection{Content moderation on e-commerce social media}
Content moderation plays an important role in social networks, reducing harmful and inappropriate content and promoting a friendly online community. Although RedNote has its content moderation mechanism to regulate illegal and harmful information, we find that there is a need for more specific guidelines for body management information. First, RedNote started as an e-commerce platform and commercial activities remain an integral part of this community. In addition to explicit advertisements, many UGCs on RedNote also focused on evaluations and discussions of products, sometimes with a blogger sponsored by the associated company. Therefore, it is important for RedNote and similar platforms to monitor and filter UGC as product information and customer reviews. Untruth or even fake product information/reviews is a common problem on e-commerce platforms \cite{he2022market}. Detecting and regulating this type of information is important for RedNote to strive as a popular social network that shares content and also maintains its commercial features.

In addition, the credibility of health information is another issue on social networks \cite{viviani2017credibility}, and our participants have expressed the need to discern healthy and scientific posts on body management. Therefore, social media platforms should incorporate mechanisms to warn about potentially harmful body management advice. Many exercise/diet routines that focus exclusively on weight reduction could have a negative impact on body health. Platforms should try to detect low-quality content and also warn users about the health implications of following social media posts. The enhanced content moderation will not only improve the user experience by providing better content but also mitigate the risks of health misinformation. 

\subsubsection{Support body management content creators}
Improving the design of the user experience is crucial to facilitating the search for information on body management. However, as our findings suggest, the factors that affect user decisions are often highly related to the characteristics of the blogger. Therefore, we believe that a significant impact can be made through design implementations that empower content creators. By providing them with the right tools and support, we can encourage the production of high-quality, user-friendly posts that not only benefit the end-users but also contribute to a more engaging and informative community.

First, the authenticity of the content emerged as a decisive factor in user trust in body management posts. The platform could help bloggers improve authenticity by allowing them to verify their identity and upload their professional certificates. In addition, since photo/video editing becomes a concern for weight loss results, platforms may verify and label editing-free content \cite{pandey2023overview} to allow bloggers to demonstrate authenticity. Moreover, as different media formats have unique strengths for presenting body management tips, the platform can support video bloggers through the auto-generation of text and graphical versions of their videos, catering to users who prefer concise and easily accessible information.

\section{Limitations}
There are several limitations that should be considered when interpreting the findings of this study. First, all participants were young, well-educated women. Although this sample reflects the predominant demographic of RedNote users, it does not represent the full diversity of social media users engaging in body management practices. Future work could include a broader range of age groups, genders, and educational backgrounds to capture more varied information behaviors and attitudes toward body management. Second, the study focused exclusively on RedNote, a female-dominant, visually oriented, and commerce-driven platform. Some findings—such as the emphasis on influencer credibility, product promotion, and visual presentation of body management outcomes—are likely platform-specific, shaped by RedNote’s combination of social networking, short-form video, and e-commerce affordances. Extending the study to other platforms could help determine whether similar evaluation practices emerge in different sociotechnical environments. Third, because RedNote is primarily used by Chinese users, the findings are deeply embedded in the cultural context of Chinese beauty and health norms, which emphasize slenderness, self-discipline, and the moral value of effortful self-improvement. These ideals shape how users interpret and act upon body management information. Comparative studies across cultural settings could clarify which aspects of users’ evaluative practices are culture-specific and which may reflect broader, cross-cultural patterns of social media–mediated body management. Lastly, as an exploratory qualitative study, our work identifies key factors that influence users’ perceptions and practices related to body management information. Future research should further examine how these factors interact over time and across different platform cultures, helping to refine and generalize the theoretical insights derived from this study.

\section{Conclusion}
Digital technologies such as social networks have significantly changed the way we seek information, affecting our lifestyle choices. This study explored how users interact with body management information on RedNote, a social-commerce platform predominantly used by women in China. We observed that users are drawn to body management content that aligns with a healthy lifestyle instead of focusing on weight reduction. Their perception of posts is shaped not just by the quality and presentation of the information but also by complex social factors, such as the influence of popular bloggers and their relatability to their real-life experiences. However, challenges arise as users can find it difficult to discern credible, healthy body management advice and to follow through with suggestions from social media. The findings emphasize the need for design improvements on platforms like RedNote to better support users in accessing reliable health information and to assist content creators in addressing user needs. As social networks become an important source of information about body management, it is essential to navigate them with an informed approach to ensure the health and welfare of users.

\balance
\bibliographystyle{ACM-Reference-Format}
\bibliography{sample-base, LRB}


\end{document}